%% file: p4-codel.tex
\documentclass[conference]{IEEEtran}
\IEEEoverridecommandlockouts
\usepackage{cite}
\usepackage{listings}
\usepackage{amsmath,amssymb,amsfonts}
\usepackage{algorithmic}
\usepackage{graphicx}
\usepackage{subcaption}
\usepackage{textcomp}
\usepackage{xcolor}
\usepackage{todonotes}
\def\BibTeX{{\rm B\kern-.05em{\sc i\kern-.025em b}\kern-.08em
    T\kern-.1667em\lower.7ex\hbox{E}\kern-.125emX}}
\usepackage{csquotes}
\usepackage{dblfloatfix}    

\definecolor{mygreen}{rgb}{0,0.4,0}
\definecolor{mygray}{rgb}{0.5,0.5,0.5}
\definecolor{mymauve}{rgb}{0.58,0,0.82}

\usepackage[font=footnotesize]{caption}
\DeclareCaptionFormat{listing} {
   \parbox{\textwidth}{\hspace{-0.2cm}#1#2#3}
}

\newboolean{komversion}
\setboolean{komversion}{false} 

\ifthenelse{\boolean{komversion}}{
    \usepackage{eso-pic}
    \AddToShipoutPictureBG{
        \AtPageUpperLeft{%
            \raisebox{-3\baselineskip}{
				\makebox[\paperwidth]{\colorbox{yellow!40}{\begin{minipage}{0.85\paperwidth}\centering\small
                     {Ralf Kundel, Amr Rizk, Jeremias Blendin, Boris Koldehofe, Rhaban Hark, Ralf Steinmetz.
						\emph{P4-CoDel: Experiences on Programmable Data Plane Hardware.} To appear in the Proceedings of IEEE International Conference on Communications, IEEE, 2021. }

                        \end{minipage}}}}%

                    }

                    \AtPageLowerLeft{%

                        \raisebox{3\baselineskip}{

\makebox[\paperwidth]{\fbox{\begin{minipage}{0.85\paperwidth}\scriptsize

                                        {The documents distributed by this server have been provided by the contributing authors as a means to ensure timely dissemination of scholarly and technical work on a non-commercial basis. Copyright and all rights therein are maintained by the authors or by other copyright holders, not withstanding that they have offered their works here electronically. It is understood that all persons copying this information will adhere to the terms and constraints invoked by each author's copyright. These works may not be reposted without the explicit permission of the copyright holder.}

                                    \end{minipage}}}}%

                    }

    }

}{}

\begin{document}

\title{P4-CoDel: Experiences on Programmable Data Plane Hardware\\
}

\author{
\IEEEauthorblockN{
Ralf Kundel\IEEEauthorrefmark{1},
Amr Rizk\IEEEauthorrefmark{5},
Jeremias Blendin\IEEEauthorrefmark{3},
Boris Koldehofe\IEEEauthorrefmark{4},
Rhaban Hark\IEEEauthorrefmark{1},
Ralf Steinmetz\IEEEauthorrefmark{1}}

\IEEEauthorblockA{\IEEEauthorrefmark{1}
Multimedia Communications Lab,
\textit{Technical University of Darmstadt, Germany}\\
\{ralf.kundel, rhaban.hark, ralf.steinmetz\}@kom.tu-darmstadt.de}

\IEEEauthorblockA{\IEEEauthorrefmark{5}
\textit{University of Ulm, Germany}\\
amr.rizk@uni-ulm.de}

\IEEEauthorblockA{\IEEEauthorrefmark{3}
Barefoot Networks, an Intel Company,
\textit{CA, US}\\
jeremias.blendin@intel.com}

\IEEEauthorblockA{\IEEEauthorrefmark{4}
\textit{University of Groningen, Netherlands}\\
b.koldehofe@rug.nl}
}
\maketitle

\begin{abstract}
Fixed buffer sizing in computer networks, especially the Internet, is a compromise between latency and bandwidth.
A decision in favor of high bandwidth, implying larger buffers, subordinates the latency as a consequence of constantly filled buffers.
This phenomenon is called \textit{Bufferbloat}.
Active Queue Management (AQM) algorithms such as CoDel or PIE, designed for the use on software based hosts, offer a flow agnostic remedy to Bufferbloat by controlling the queue filling and hence the latency through subtle packet drops.

In previous work, we have shown that the data plane programming language P4 is powerful enough to implement the CoDel algorithm.
While legacy software algorithms can be easily compiled onto almost any processing architecture, this is not generally true for AQM on programmable data plane hardware, i.e., programmable packet processors.
In this work, we highlight corresponding challenges, demonstrate how to tackle them, and provide techniques enabling the implementation of such AQM algorithms on different high speed P4-programmable data plane hardware targets.
In addition, we provide measurement results created on different P4-programmable data plane targets.
The resulting latency measurements reveal the feasibility and the constraints to be considered to perform Active Queue Management within these devices.
Finally, we release the source code and instructions to reproduce the results in this paper as open source to the research community.
\end{abstract}


\section{Introduction}

\input{chapter/Introduction}

\section{The CoDel AQM Algorithm}
\input{chapter/problemgraph}

\section{Hardware Implementation}
\input{chapter/design}

\section{Evaluation}
\input{chapter/evaluation}

\section{Related Work}
\input{chapter/relatedwork_new}

\section{Conclusion}
\input{chapter/conclusion}

\section*{Acknowledgment}

This work has been supported by Deutsche Telekom through the Dynamic Networks 8 project, and in parts by the German Research Foundation (DFG) as part of the projects B4 and C2 within the Collaborative Research Center (CRC) 1053 MAKI as well as the DFG project SPINE.
Furthermore, we thank our colleagues and reviewers for their valuable input and feedback.

\bibliographystyle{IEEEtran}
\bibliography{bibliography}

\end{document}

%% file: chapter/Introduction.tex
\emph{Bufferbloat} describes a phenomenon of high latencies observed in networks configured with large static buffer sizes~\cite{gettys2012bufferbloat}.
For a single TCP traffic flow, it is known that the buffer size that maximizes the TCP throughput is directly proportional to the round trip time (RTT).
This is also true for the case of multiplexing many homogeneous flows having the same RTT except of a correction prefactor~\cite{appenzeller2004sizing}.
As traffic flows usually have widely different RTTs and throughput maximization is not the main goal for many contemporary networking applications, but rather latency  minimization, an idea of controlling flow delays through subtle packet drops is experiencing a renaissance~\cite{Floyd1993RED}.
The algorithmic version of this idea is denoted Active Queue Management (AQM) and is based on the sensitive reaction of the transport protocol congestion control, typically a TCP variant, to packet drops.
By dropping packets earlier than at a full buffer, the sender congestion control algorithm receives an \emph{early signal} to reduce the sending data rate.
In turn, this leads to the buffer filling and, hence, the \emph{queueing delay} to remain relatively small.
In recent years, two stateful and self-tuning AQM algorithms, i.e., CoDel~(RFC~8289)\cite{rfc8289} and PIE~(RFC~8033), have been presented and widely adopted. 
In addition to those two approaches, there exist many variants of AQM algorithms with different goals and assumptions~\cite{sivaraman2013no}.

\begin{figure}[t!]
	\centering
	\includegraphics[width=1.0\linewidth]{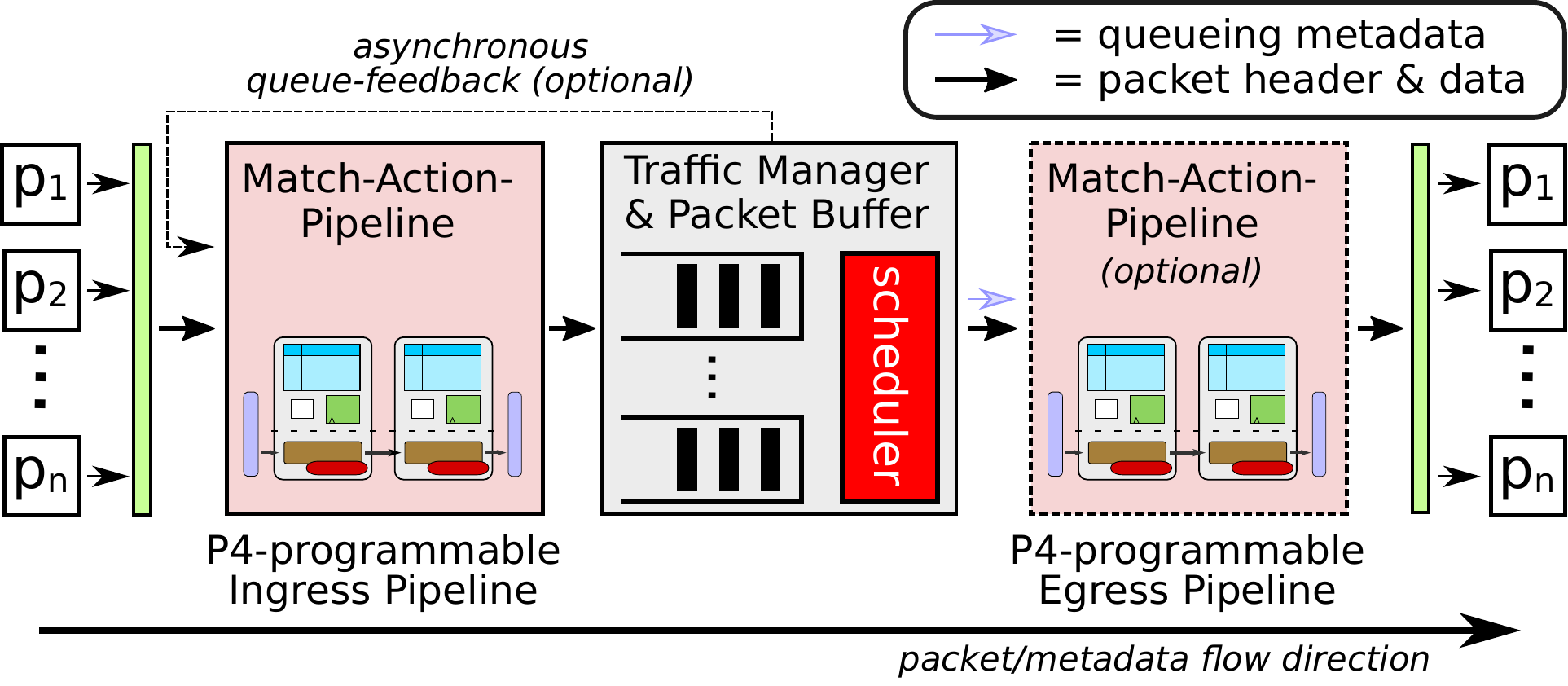}
	\caption{
	 Functional building blocks of P4-programmable ASICs. The light-red parts are P4 programmable. Packet queueing is not part of P4 and its configuration is vendor dependent. A programmable Match-Action Pipeline after the packet buffer is not given in all currently existing architectures.
	}
	\label{fig:p4-pipe_queue}
\vspace{-1.em}
\end{figure}

In a previous work, we have shown that such AQM algorithms can be expressed with programming languages aimed at controlling network switch data plane behavior~\cite{Kundel2018Codel}.
However, it remained open how feasible it is to realize such AQM algorithms on packet processors, as these algorithms were not conceptualized to run on data plane network hardware but rather on software based consumer edge devices.
Indeed, for many networking applications packet processing with high throughput is required and can only be ensured by algorithms realized directly within the data plane.
Further, it remained open how different programmable networking hardware affect the algorithms performance and which additional challenges arise in different architectures.

To understand the problem of running AQM algorithms on programmable data plane hardware, a deeper look into the pipeline of modern packet processors is required.
 The internal pipeline of packet processors, including network switches, is typically similar to the architecture depicted in Figure~\ref{fig:p4-pipe_queue}.
Packets enter on one of the $n$ ports on the left side and are multiplexed into a single match-action-pipeline.
Within this pipeline operations, e.g. lookups, on packet header fields and metadata can be performed.
After that, the traffic manager is responsible for packet queueing and scheduling.
Optionally, depending on the architecture, a second match-action-pipeline can be applied on the packets before demultiplexing and sending the packets out on the specified egress port.
Note that packets can flow only from left to right and \emph{algorithmic loops within the pipeline are not possible}.
By that, a high and deterministic processing performance can be guaranteed which is crucial for network functions running at line rate~\cite{kundel2020bng}, including AQM algorithms running within the network.

In case of programmable packet processors built on top of the Protocol Independent Switch Architecture~(PISA)~\cite{bosshart2013forwarding},  these ingress and egress match-action-pipelines are programmable within limitations.
The programming language P4~\cite{bosshart2014p4} represents the currently most widely accepted approach of programming the depicted ingress and egress match-action-pipelines.
However, the configuration of the packet buffer in the middle of the switch architecture (see Figure~\ref{fig:p4-pipe_queue}) as well as the specific algorithms used by this engine for queueing, scheduling and buffer management are out of the scope and purpose of this language.
Nevertheless, useful metadata produced inside the Traffic manager, such as the packet queueing delay, can be passed alongside with other packet metadata into the egress pipeline.
Alternatively, depending on the actual architecture, the current buffer utilization can be passed asynchronously to the ingress pipeline.

The work at hand focuses on the feasibility of realizing AQM algorithms on programmable hardware devices, which creates a variety of challenges compared to the implementations on classical commodity processor.
Here, we analyze the required alterations of an established stateful AQM algorithm (CoDel) that make it possible to implement it.
Note that our key findings are not dependent on the specific AQM algorithm at hand.
We evaluate P4-Codel for P4-NetFPGAs~\cite{ibanez2019p4}, Netronome SmartNICs and Intel Tofino switches.

The main contributions of this paper are:
\begin{itemize}
\item The analysis of CoDel's AQM implementation control flow, particularly considering information dependencies,
\item how to transform an algorithm designed for CPUs for packet processing pipelines and a detailed implementation description of CoDel for existing P4~hardware,
\item evaluation results of characteristic properties for these different hardware implementations,
\item an open source implementation of the presented algorithm for two different hardware architectures including reproduction instructions.
\end{itemize}

%% file: chapter/problemgraph.tex
\begin{figure*}[!t]
\vspace{0.05em}
	\centering
	\includegraphics[width=1.0\linewidth]{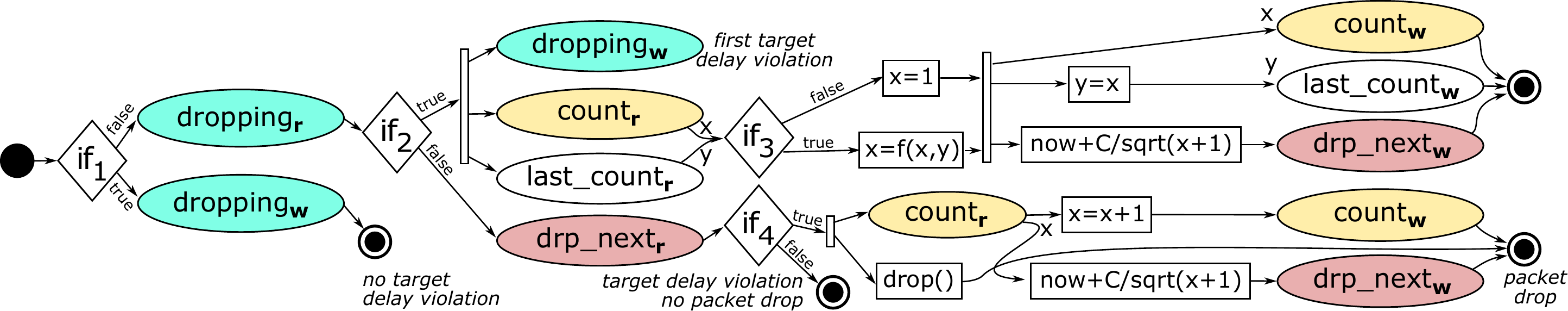}
	\caption{General stateful data centric program flow of the CoDel-algorithm. Read access on stateful variables is indicated by \textbf{r}, write operations by \textbf{w}.}
	\label{fig:dependency-graph}
\end{figure*}

In the following, we use the CoDel AQM-algorithm presentation given in Listing~\ref{lst:codel} to illustrate the stateful data dependencies within the algorithm.
The effectiveness of CoDel-algorithm, presented in 2018 by Nichols \textit{et al}~\cite{rfc8289}, will not be discussed further.
Figure~\ref{fig:dependency-graph} shows the control flow graph of the algorithm that includes four stateful variables: 1)~\textit{dropping}, 2)~\textit{count}, 3)~\textit{last\_count} and 4)~\textit{drp\_next}.
The value of a stateful variable persist the processing time of a single packet.

The algorithm can be regarded as state machine consisting of two states: 1) \textit{dropping} and 2) \textit{not dropping}.
As soon as the observed queueing delay exceeds the \textit{TARGET} of 5~ms, the state of the state machine is changed to \textit{dropping} which, however, does not imply an immediate packet drop~(\textit{if\_2}).
After waiting for the time \textit{INTERVAL} until time \textit{drp\_next} within the \textit{dropping} state, the first following packet is dropped and the \textit{counter} of dropped packets is increased~(\textit{if\_4}).
From this point on, the interval between dropping packets is continuously decreased as $\frac{INTERVAL}{\sqrt{count}}$ until the queueing delay falls below the \textit{TARGET} delay.
Then, the state changes back to \textit{not dropping}~(\textit{if\_1}).
In case of a recently left \textit{dropping} state, the new dropping rate is initialized higher than 1~(\textit{if\_3}).

\begin{figure}[b!]
\vspace{-0.5em}
\begin{lstlisting}[caption={CoDel Pseudocode, based on RFC 8289~\cite{rfc8289}.}, xleftmargin=0.85em, label=lst:codel, language=C]
#define TARGET 5 //ms
#define INTERVAL 100 //ms
Queue q; State s;

upon receive packet p:
 //no target delay violation? (if_1)
 if(p.queue_delay < TARGET || q.byte < IFACE_MTU):
   s.dropping = false
   continue
 //first packet which violates delay target? (if_2)
 if(s.dropping == false):
   s.dropping = true
   tmp = s.count
   //drop frequency steady state? (if_3)
   if((s.count - s.last_count > 1) &&
       (now - s.drp_next_packet < 16*INTERVAL)):
     //optimization for faster packet dropping
     s.count = s.count - s.last_count
   else:
     s.count = 1
   s.drp_next_packet = now + INTERVAL/sqrt(s.count)
   s.last_count = tmp
   continue
 //drop state for at least x time units? (if_4)
 if(s.dropping && s.drp_next_packet <= now):
   p.drop()
   s.count++
   s.drp_next_packet = now + INTERVAL/sqrt(s.count)
\end{lstlisting}
\end{figure}

The CoDel algorithm is conceptualized for sequential packet processing, i.e., no parallel processing of multiple packets.
Thus, the processing of $packet_n$ is expected to be completed before the processing of $packet_{n+1}$ starts.
Otherwise, e.g., the read operation on the stateful variable \textit{drp\_next} for $packet_{n+1}$ is performed before the write operation on this variable by $packet_n$ is completed and by that unexpected behavior occurs.

Note that from an algorithmic perspective a partial overlapping processing of multiple packets is possible.
If for each stateful variable the write operation is performed before the read operation of the subsequent packet, the algorithm is executed correctly.
In the concrete case of CoDel, operations on the stateful \textit{dropping} variable can be considered independently to the other three state variables and by that executed in parallel for consecutive packets.
As noticeable from  Figure \ref{fig:dependency-graph}, the other three stateful variables cannot be isolated as the operations depend on each other.
Thus, all operations on these three variables must be executed as an atomic block.

\begin{figure}[t!]
	\centering
	\includegraphics[width=1.0\linewidth]{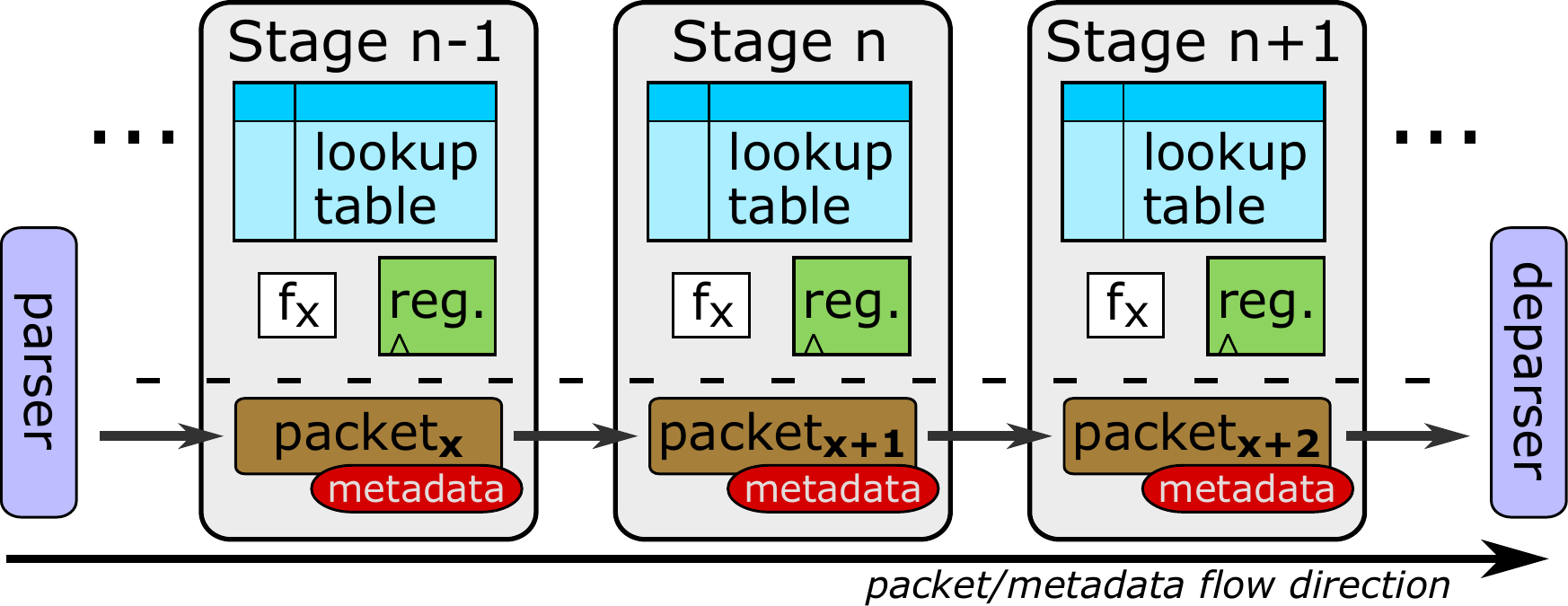}
	\caption{Generic acyclic P4-programmable Match-Action Pipeline including packet (de)parsing.
	Periodically, after each pipeline clock cycle, all packets are forwarded by one stage.
	Backwards information flow is not possible.
	Each stateful information is attached to a register in one dedicated stage.
	}
	\label{fig:p4-pipe}
\vspace{-1em}
\end{figure}

The structure of programmable packet pipelines, depicted in Figure~\ref{fig:p4-pipe}, does not allow cyclic data dependencies over multiple stages.
This means, a stateful variable of the algorithm located in a register of stage $n-1$ can only be read there and used for further computations in stage $n-1$, $n$ and $n+1$.
In this case, however, a read-compute-write cycle is only possible within stage $n-1$ as the result cannot be fed back from a later stage to stage $n-1$ which is then already processing the subsequent packet.
As a consequence, each atomic stateful operation must be executed within one single pipeline stage including reading and writing all required variables.
We observe that other AQM algorithms such as PIE can be analyzed in a similar way in order to obtain the stateful data dependencies.

%% file: chapter/design.tex
In the following, we describe three efforts to migrate and run the CoDel algorithm on significantly different P4-programmable hardware platforms.
Recall that P4 is designed to describe packet header processing and not for queueing.

\subsection{PISA-architecture switches}
The PISA platform allows the ingress and egress pipeline, consisting each of $N$ stages, to be programmed with P4.
As the queueing delay information is only available within the egress pipeline we decided to implement CoDel there.
In addition, the traffic manager needs to be configured accordingly to a constant rate, e.g., $100$ Mbit/s, in order to build up a queue.

As stated before, the algorithm must be loop- and cycle-free over different pipeline stages and information can only flow from the left to the right as shown in Figure~\ref{fig:p4-pipe}.
In order to achieve this, we adapt the algorithm as shown in Listing~\ref{lst:codel_rewrite}.
This adaptation targets the mapping of each atomic operation onto a single match-action pipeline stage.
Note that in case of Intel Tofino, following the PISA-architecture, allows the P4 program to execute a small number of different operations, such as $f_x$,  on the contents of the register within a single stage using a special ALU, called stateful ALU (see  Figure~\ref{fig:p4-pipe}).

First, we determine whether the \textit{TARGET} delay is violated or not in function~\textit{f\_1}.
This operation is stateless and can be performed for each packet independently.
Second, we update the \textit{dropping} state and deviate if the current delay violation is not preceded by a delay violation, which means the state is changing from \textit{non\_dropping} to \textit{dropping}.
Third, the decision whether a packet should be dropped, the computation of the next drop time (\textit{drp\_next}) and incrementing the drop \textit{counter} are performed within another  stateful ALU.
The output of this ALU, either $0$ or $1$, indicates if the current packet should be dropped if the \textit{TARGET} delay is violated as well (\textit{codel\_drop}).
The square root function can be approximated very well by the math unit within the stateful ALU.
Last, the final dropping decision is performed by checking if the \textit{TARGET} delay is violated (output of $f_1$) and the second stateful ALU marks this packet to be dropped.

\begin{figure}[t]
	\centering
	\includegraphics[width=0.99\linewidth]{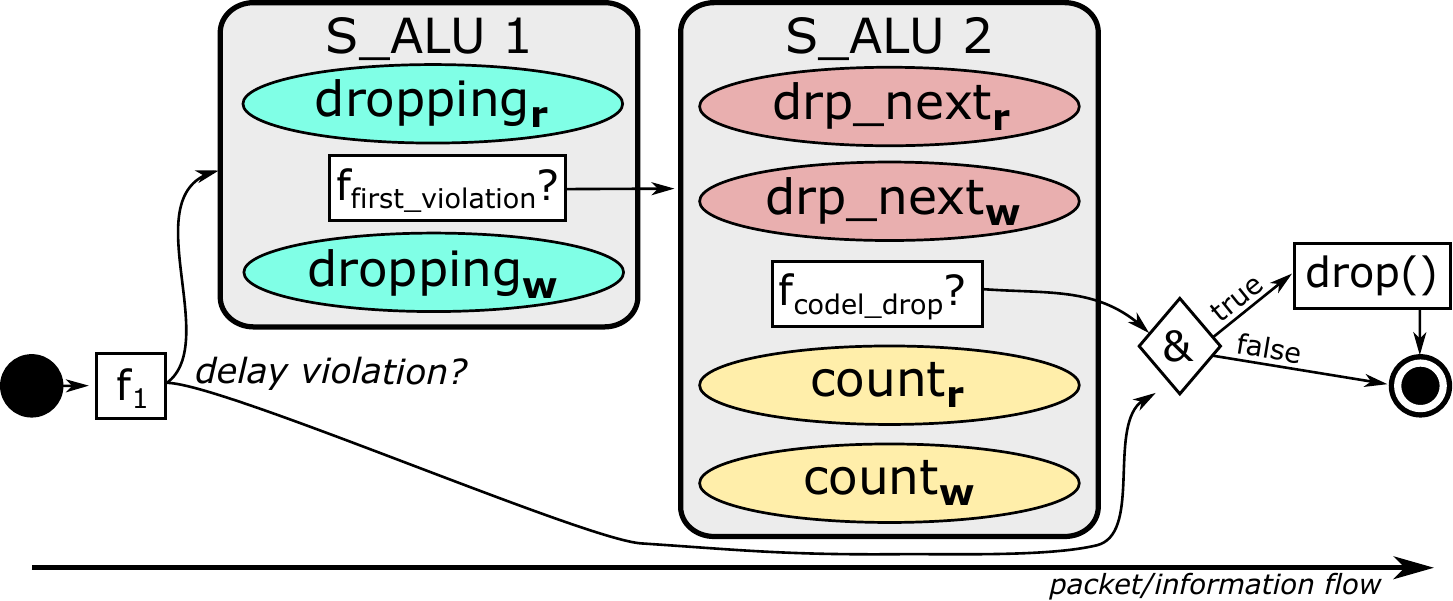}
	\caption{Stateful data centric program flow of the CoDel-algorithm rewritten without cyclic data dependencies optimized for the PISA architecture. Each gray box represents a stateful ALU which guarantees atomic execution within one single pipeline stage.}
	\label{fig:dependency-graph_rewrite}
	\vspace{-1em}
\end{figure}

The mapping of these functional blocks on the PISA pipeline is depicted in Figure~\ref{fig:dependency-graph_rewrite}.
In total, the algorithm requires four pipeline stages: 1)~performing the $f_1$ computation, 2,3) stateful ALU~1 and~2,  and 4)~the final drop decision.

Compared to the algorithm in Listing~\ref{lst:codel}, the optimization for more frequent packet dropping in the beginning which is marked as \textit{if\_3} could not be implemented on this PISA hardware.
The reason for that is, that the number of registers and arithmetic operations 
that can be performed of them by a single stateful ALU is limited in order to provide real time guarantees and high performance up to the line rate of $100 Gbit/s$ per port for many ports in parallel.
Furthermore, for newer hardware generations we expect even more flexible stateful ALUs and, as shown later in evaluation section, the impact of this limitation on the performance is very limited.

\noindent\textbf{Fq\_CoDel:} An evolution of CoDel is given by fq\_CoDel which can be realized on PISA platforms as well.
This algorithm distributes incoming packets in the ingress pipeline over multiple queues for one single egress port.
These queues are served by a (weighted) round robin scheduler and for each queue a CoDel state is required in the egress pipeline.
As the stateful ALU provide an addressable register set instead of a single register, this can be realized as well.

\begin{figure}[b!]
\vspace{-1em}
\begin{lstlisting}[caption={Rewritten CoDel algorithm for fitting into loop- and -free P4-programmable PISA pipelines. This code does not contain all implementation details and just give, together with Figure \ref{fig:dependency-graph_rewrite} an overview.}, label=lst:codel_rewrite, language=C]
#define TARGET 5 //ms
#define INTERVAL 100 //ms
Packet p; Queue q;
//check for target delay violation? (f_1)
if(p.queue_delay < TARGET || q.byte < IFACE_MTU):
  delay_violation = false
else:
  delay_violation = true
first_violation = S_ALU1.exec(delay_violation)
codel_drop = S_ALU2.exec(first_violation)
if(delay_violation && codel_drop):
  p.drop()
\end{lstlisting}
\end{figure}

\subsection{NPU-based packet processors}
\label{subsec:NPU}
The internal structure of so called Network Processing Units (NPU) is similar to a many-core processor or GPU.
Incoming packets are divided over a set of processors, that execute in parallel the compiled program code of the P4-programmable ingress pipeline.
After that, the packets are enqueued in per port egress queues with a certain rate limit, e.g., $100$ Mbit/s.
After these queues, a processing by a P4 programmable block is not possible any more.
However, as shown in Figure~\ref{fig:p4-pipe_queue}, an asynchronous feedback from the packet queues to the P4-programmable ingress pipeline is possible.
Thus, the AQM has to be implemented within the ingress pipeline and the input to the algorithm is the current delay of the queue and not the queueing delay of the current packet which are only subtly different.
We noticed that access to this information on the queue level is only possible within a microC-sandbox, executed as external P4 function, and not directly in P4.
Due to further complexities while handing over data from and to this sandbox, we decided to realize the CoDel AQM as a fixed P4-external function within a sandbox.

The approximation of the square root function was realized by an exact lookup table for the values $count=1...16$ and by an approximation for all other values.
This approximation, already presented in~\cite{Kundel2018Codel}, uses a longest prefix matcher in order to count the leading zeros of the stateful \textit{count} variable.

On this hardware architecture we observed two main issues: First, as multiple parallel processing units are running within the SmartNIC, race conditions between multiple accesses on stateful variables can occur.
Due to the internal processor architecture, there is no trivial solution to this challenge. 
Second, to the best of our knowledge, the queue depth can only be measured by the number of enqueued packets which is, in case of heterogeneous packet sizes, a major constraint.

\subsection{P4-NetFPGA}
Lastly, we investigated the P4-NetFPGA architecture~\cite{ibanez2019p4}.
This architecture provides only tiny queues per egress ports.
In addition, these queues are not rate limited and always send packets out with the link rate, i.e., in this case $10 Gbit/s$.
However, we analyzed possibilities of getting CoDel to work within the P4 pipeline.
As the stateful operations turned out to be challenging in this context, we decided to provide the AQM algorithm as a P4 external function to the pipeline, similar to the NPU-based SmartNIC approach given above in Sect.~\ref{subsec:NPU}.
Such an external AQM algorithm, however, has to be designed in a low level programming language as Verilog or VHDL.
Finally, our running CoDel-prototype for P4-NetFPGA is avoiding all components of the P4-pipeline and all AQM-crucial components are realized as standalone modules.
Consequently, our implementation would run even without a P4 framework and therefore we do not show results for P4-NetFPGA in the following.

%% file: chapter/evaluation.tex
\begin{figure}[t!]
	\centering
	\includegraphics[width=1.0\linewidth]{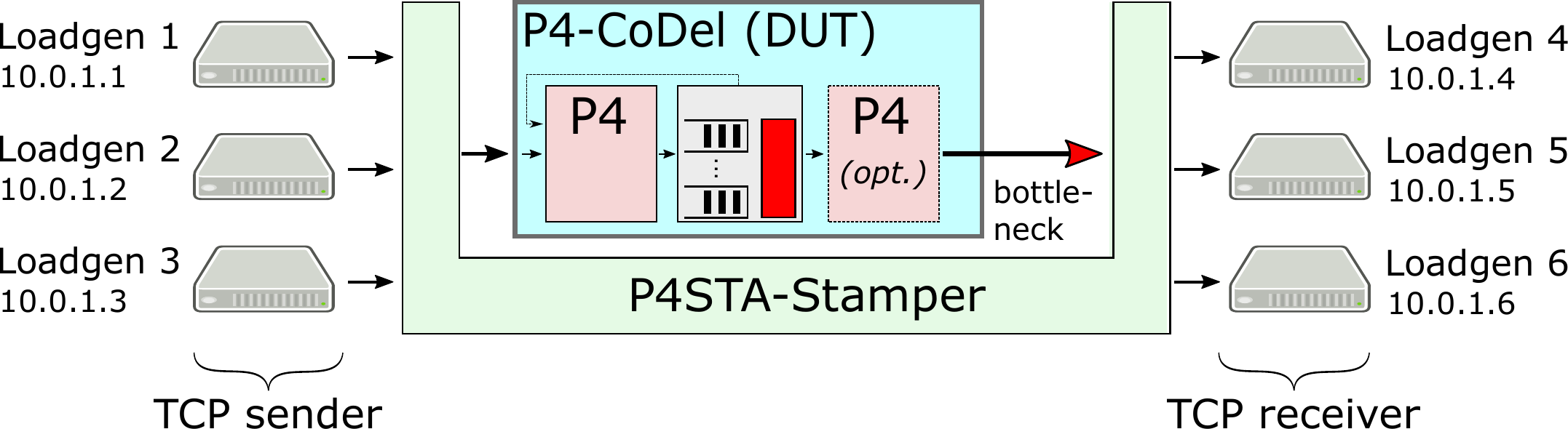}
	\caption{Data plane testbed setup based on the P4STA load aggregation framework~\cite{kundel2020p4sta}. Each server can create multiple parallel TCP flows. Each server can be assigned an additional link latency for heterogeneous RTTs. }
	\label{fig:eval_topo}
	\vspace{-1em}
\end{figure}

The following evaluation experiments are performed using a testbed built upon the P4STA load aggregation and measurement framework~\cite{kundel2020p4sta} as depicted in Figure~\ref{fig:eval_topo}.
The Device Under Test (DUT), encircled by the P4STA-Stamper for high accuracy time measurements and loss detection, represents the CoDel implementation on a P4-pipeline realized by \emph{(i)} the Intel Tofino ASIC, \emph{(ii)} Netronome NPU-based SmartNICs and the \emph{(iii)} Linux kernel as reference implementation.
All scripts for reproducing the results are available as open source software together with the CoDel source code~\footnote{https://github.com/ralfkundel/p4-codel}.

Considering Figure~\ref{fig:eval_topo} each packet sent by any of the three TCP senders on the left side is timestamped and counted before entering the CoDel implementation of the DUT.
The same is performed for each packet leaving on the right side to the receivers and by that very accurate queueing delay measurements are be performed.
TCP acknowledgments traverse the setup back smoothly without any packet loss.

\begin{figure*}[h!]
\begin{subfigure}{0.47\textwidth}
	\centering
	\includegraphics[width=\linewidth]{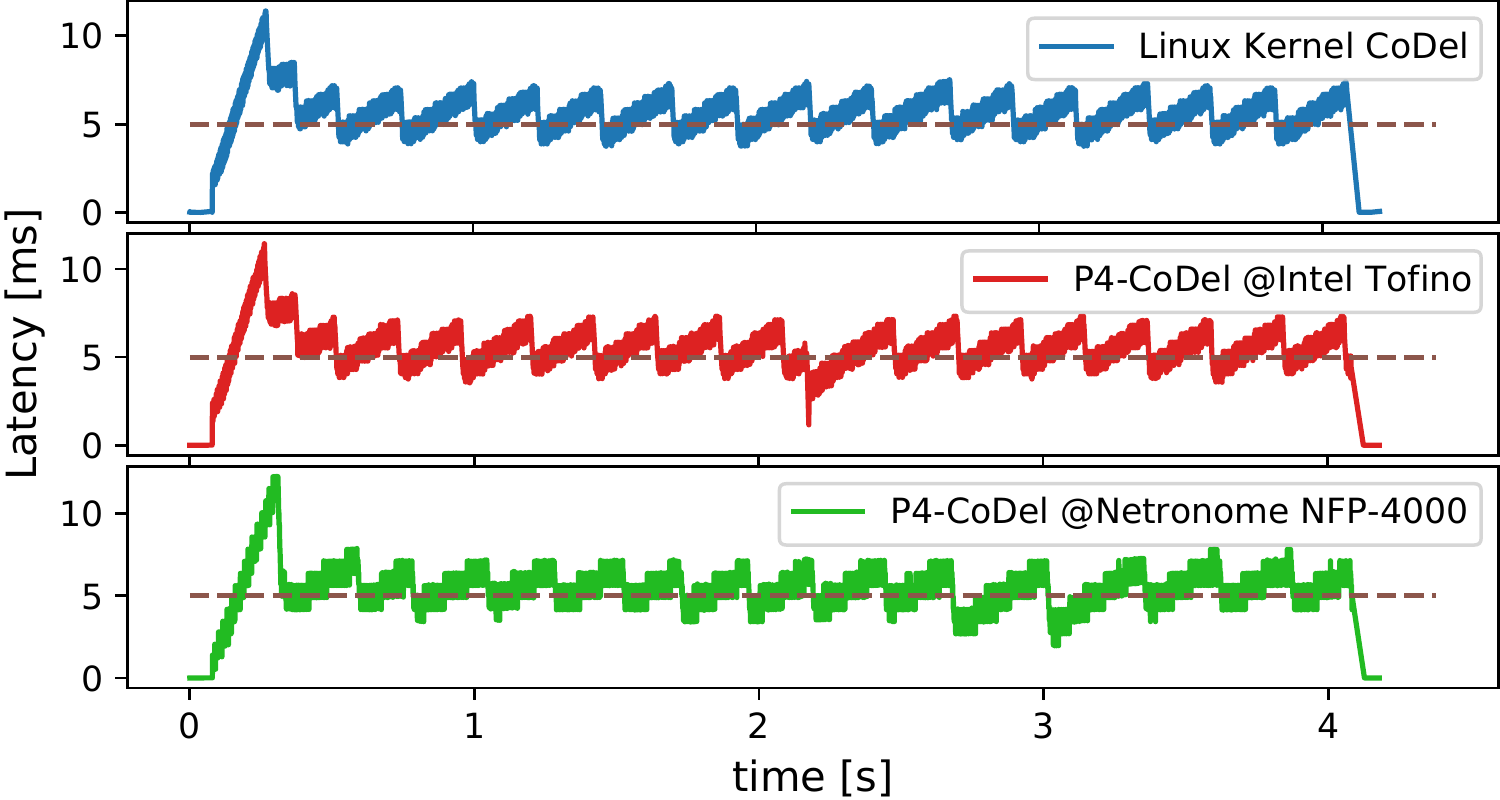}
	\caption{}
	\label{fig:eval_graph1}
	\end{subfigure}
\begin{subfigure}{0.47\textwidth}
	\centering
	\includegraphics[width=\linewidth]{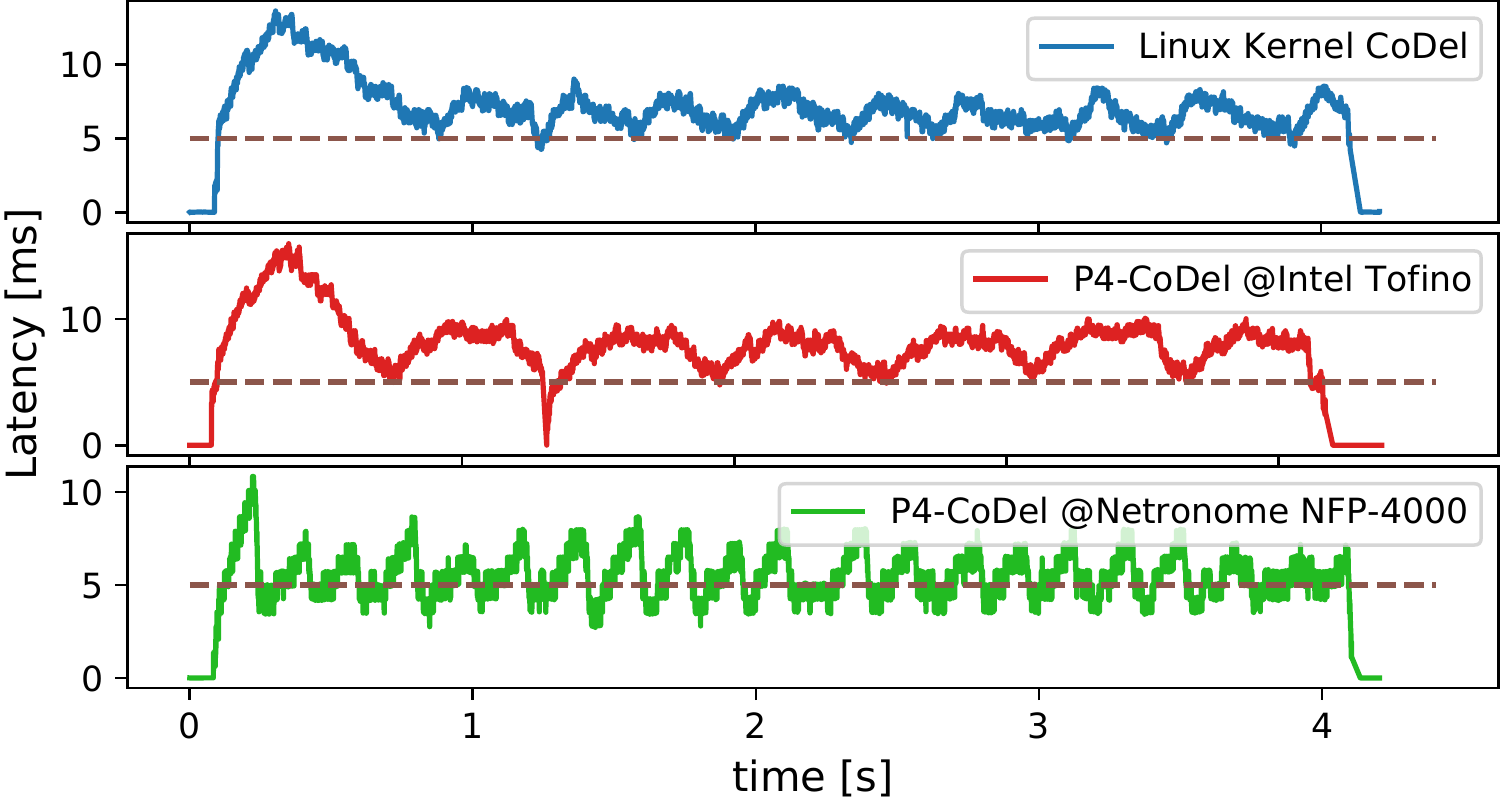}
	\caption{}
	\label{fig:eval_graph2}
\end{subfigure}
\caption{Measured latency for CoDel realizations on P4-programmable packet switching ASICs, NPU-based P4-SmartNICs, vs the Linux kernel reference implementation for (a) one and (b) three parallel TCP flows (cf. Fig.~\ref{fig:eval_topo} for measurement setup).}
	\vspace{-1em}
\end{figure*}

In a first run, a single TCP sender and receiver, using iPerf3, transfer as much data as possible over the bottleneck link.
The results for the three investigated CoDel AQM DUTs are shown in Figure~\ref{fig:eval_graph1}.
In all three scenarios an initial burst of packets is filling the buffer to a latency slightly above $10 ms$ until CoDel reacts and reduces the latency by subtle packet drops which results in a reduced TCP sending rate.
By that, the queueing delay falls periodically below the configured \textit{TARGET} delay of $5ms$.
On the rising edge one may notice a oscillation of the latency.
This behavior can be explained by a microburstiness of the TCP sender\cite{kundel2020microbursts}.
In Figure~\ref{fig:eval_graph2} we consider three senders in parallel where this patterns becomes noisy.
For the NPU-based Netronome SmartNICs this noise is slightly higher; this could be caused by the internal many-core processor architecture, however, we could neither confirm nor refute this.

In addition to considering only one single TCP flow, we will focus next on the case of multiplexing multiple flows.
Table~\ref{tbl:eval_drops} shows the number of dropped packets by the CoDel implementation and the average latency for 1,2 and 3 parallel TCP flows.
First, we notice that the number of dropped packets is strongly increasing with the number of parallel TCP flows.
This is caused by the fact that CoDel drops only one single packet and by that the corresponding TCP flow is reducing its sending rate but the other flows benefit from the released free link capacity.
As a consequence, an AQM in general has to drop more packets in order to control all flows and by that the latency.
Second, the average latency increase on the P4-ASIC when going from 1 to 3 TCP flows is significant.
This can be explained by the missing optimization (\textit{if\_3}) for this P4 target.

In Figure~\ref{fig:eval_graph2}  we show the latency for the three TCP flows where we observe a similar behavior for the P4-ASIC and the Linux kernel.
The results for the NPU-based SmartNICs suggest a TCP flow synchronization which is usually avoided by the CoDel algorithm.
Lastly note that the drop rate of the Linux kernel is slightly lower.
The reason behind this could be the difference between the kernel implementation and the RFC implementation we build upon.

\begin{figure}[t!]
	\centering
	\includegraphics[width=1.0\linewidth]{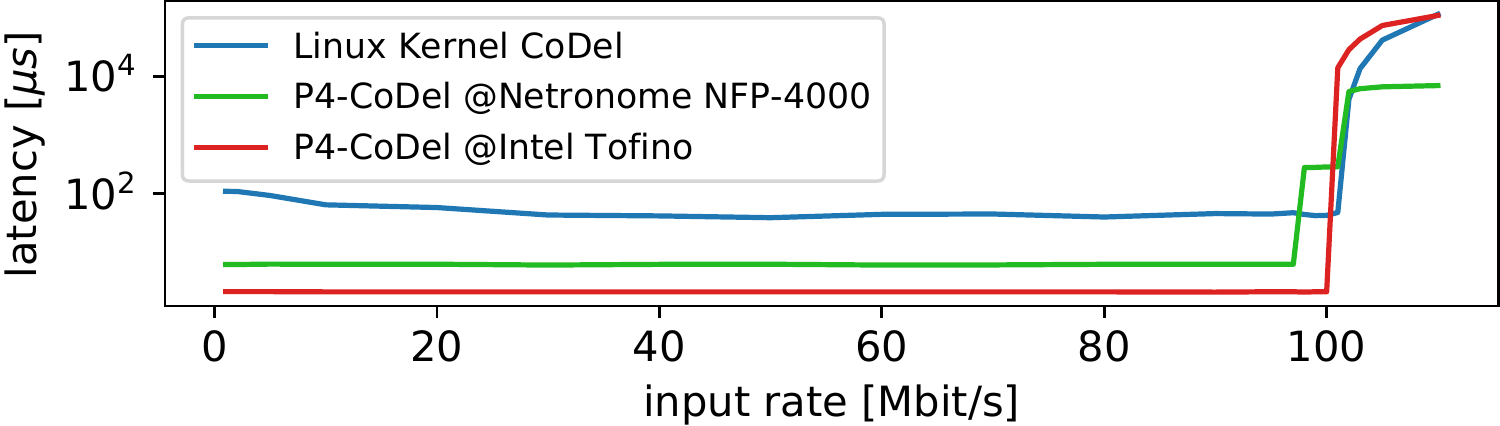}
	\caption{Average latency for isochronous shaped UDP traffic (no congestion control) with a varying rate and a CoDel rate limit of $100 Mbit/s$. }
	\label{fig:rate_sweep}
	\vspace{-1.3em}
\end{figure}

Figure~\ref{fig:rate_sweep} depicts the observed average latency (mainly due to CoDel processing on different hardware) for a rate scan with shaped constant rate UDP streams. 
Note that the CoDel shaping rate is $100 Mbit/s$, i.e.  the latency is expected to start increasing around that point due to queue filling within the DUT.
The Linux implementations shows an increased average latency for small data rates due to the Linux kernel interrupt behavior.
The latency of Netronome SmartNICs starts increasing at $97 Mbit/s$, however, a persistent queue building starts as expected at rate higher than the CoDel shaping rate.

\begin{table}[b!]
\center
\begin{tabular}{c|ll|ll|ll}
  & \multicolumn{2}{c}{Linux} & \multicolumn{2}{c}{P4-NPU} & \multicolumn{2}{c}{P4-ASIC} \\
\#flows  & loss     & latency   & loss      & latency   & loss       & latency   \\
  \hline
1 & 0.05\%   & 5.7 ms         & 0.17 \%   & 5.51 ms        & 0.16 \%    & 5.59 ms        \\
2 & 0.17\%   & 6.5 ms         & 0.24 \%   & 6.25 ms        & 0.36 \%    & 6.86 ms        \\
3 & 0.33\%   & 7.2 ms         & 0.45 \%   & 6.42 ms        & 0.44 \%    & 8.17 ms
\end{tabular}
\caption{Observed packet drops and latency for the different investigated targets. Each run is $4 s$ with a rate of $100 Mbit/s$ and $~33*10^3$ packets.}
\label{tbl:eval_drops}
\end{table}

%% file: chapter/relatedwork_new.tex
Programmable queueing and scheduling algorithms have been discussed, e.g., in Sharma \textit{et al} \cite{sharma2018approximating} who proposed in 2018 a queueing mechanism, called ``Approximate Fair Queueing'', prioritizing packets in the programmable data plane in order to achieve shorter flow completion times.
In a follow up work they proposed the idea of programmable queues based on the construct of calendar queues~\cite{sharma2020CalendarQueues} which provide high flexibility in programming schedulers.
However, this approach relies on queues which can be re-parameterized from within the data plane which is not supported by existing switches.
Another approach of programming queues within switches is the approximation of Push-In-First-Out~(PIFO) queues by conventional FIFO queues.
Alcoz \textit{et al}~\cite{alcoz2020sppifo} presented a P4-based implementation, called ``SP-PIFO'', of this approach running in existing data planes with a behavior that is very close to ideal PIFO queues.
The authors of ``ConQuest''~\cite{chen2019FineGrained} tackled the impact of short-lived packet surges by monitoring the network traffic from within the data plane and apply some queue management.
Specifically, similar to our work, they present a prototype that is based on a P4-programmable Intel Tofino switch which is able to identify these surges and avoid congestion by early marking or dropping certain packets.

In a previous work we have shown the feasibility of CoDel-AQM algorithm in the programming language P4~\cite{Kundel2018Codel}.
Based on that, Papagianni \textit{et al} have introduced ``PI2 for P4''~\cite{Papagianni2019PI2} that provides a P4 implementation of the AQM algorithm PI2~\cite{Schepper2016PI2} for programmable data planes.

%% file: chapter/conclusion.tex
Active Queue Management (AQM) algorithms enable high throughput and low latency simultaneously and provide a practical solution to the \emph{Bufferbloat} phenomenon.
In this work, we have shown how recent self-tuning AQM algorithms can be implemented on P4-programmable hardware without built-in queue management capabilities achieving much better performance than in software systems.
Further, we discussed different challenges that arise depending on the chosen hardware platform to run the AQM algorithm. 
While the programming language P4 itself is quite powerful, writing compilable and correct behaving AQM algorithms leads to challenges due to hardware constraints.
With this in mind, we note that correct hardware implementations, are far superior in terms of speed and deterministic behavior to software implementations.

We investigated multiple P4 target platforms with quite diverse internal architectures:
The PISA architecture, concrete the Intel Tofino ASIC, is very restrictive on the one hand, but turned out to be the most powerful in terms of performance one on the other hand.
The investigated NPU-based SmartNICs can be programmed in a very flexible way but facilitate causing race conditions within the developed AQM-program.
Further, the P4-NetFPGA architecture does not support large packet buffers and hinders a successful implementation from an algorithmic point of view.
Only by strong modifications outside the P4-pipeline this architecture is able to perform AQM, subverting the idea of this work.

Finally, our results show that programmable data plane hardware is not only suitable for header processing but also for flexible queue management.
We noticed that advanced queue management algorithms require stateful processing capabilities in the underlying hardware.
To enable a seamless integration of such algorithms, we hope with this paper to start a discussion on AQM interface definitions for P4 programmable pipelines.